\documentclass[prl,twocolumn,floatfix,showpacs,preprintnumbers,superscriptaddress,amsmath,amssymb]{revtex4}

\usepackage{graphicx}
\usepackage{amssymb}
\usepackage{color}
\usepackage{wasysym}

\begin{document}

\title{Anisotropic magnetoresistance in ferromagnetic atomic-sized metal contacts}

\author{M. H\"afner}
\affiliation{Institut f\"ur Theoretische Festk\"orperphysik
and DFG-Center for Functional Nanostructures, Universit\"at Karlsruhe,
D-76128 Karlsruhe, Germany}
\affiliation{Departamento de F\'{\i}sica Te\'orica de la Materia
Condensada, Universidad Aut\'onoma de Madrid, E-28049 Madrid, Spain}

\author{J. K. Viljas}
\affiliation{Institut f\"ur Theoretische Festk\"orperphysik
and DFG-Center for Functional Nanostructures, Universit\"at Karlsruhe,
D-76128 Karlsruhe, Germany}
\affiliation{Forschungszentrum Karlsruhe, Institut f\"ur Nanotechnologie,
D-76021 Karlsruhe, Germany}

\author{J. C. Cuevas}
\affiliation{Departamento de F\'{\i}sica Te\'orica de la Materia
Condensada, Universidad Aut\'onoma de Madrid, E-28049 Madrid, Spain}

\date{\today}

\begin{abstract}
  Recent experiments in ferromagnetic atomic-sized contacts have shown
  that the anisotropic magnetoresistance (AMR) is greatly enhanced and
  has an asymmetric angular dependence as compared with that of bulk
  samples.  The origin of these effects is still under debate. In this
  work we present a theoretical analysis of the AMR in atomic contacts
  of the $3d$ ferromagnetic materials. Our results strongly suggest
  that the anomalous AMR stems from the reduced symmetry of the atomic
  contact geometries. We also present calculations supporting the idea
  that the pronounced voltage- and temperature dependence in some
  experiments can be attributed to impurities near the constrictions.
\end{abstract}


\maketitle
When a metallic wire is shrunk to the atomic scale, its transport
properties are significantly altered~\cite{Agrait2003}. One
interesting and novel example is the anisotropic magnetoresistance
(AMR) that arises when the magnetization throughout a ferromagnetic
device is rotated uniformly a certain angle $\theta$ with respect to
the current direction. In polycrystalline bulk samples such rotation
induces a relative change in the conductance $\Delta G / G $ that
varies as $\cos^2\theta$ with an amplitude on the order of
$1\%$~\cite{McGuire1975}. Recently Bolotin \emph{et
  al.}~\cite{Bolotin2006} found that the AMR of permalloy atomic-sized
contacts can be considerably enhanced as compared with bulk samples
and that it exhibits an angular dependence clearly deviating from
the $\cos^2\theta$ law. Additionally, they found a significant voltage
dependence on the scale of millivolts, which led them to interpret the
effect as a consequence of conductance fluctuations due to quantum
interference~\cite{Adam2006}. Independently, Viret and
coworkers~\cite{Viret2006} reported similar results in Ni contacts,
but also the occurrence of conductance jumps upon rotation of the
magnetization. Similar stepwise variations of the conductance have
been found in Co nanocontacts~\cite{Sokolov2007}.

The jumps have been interpreted as a manifestation of the so-called
ballistic AMR (BAMR)~\cite{Velev2005}. According to this theoretical
concept the rotation of the magnetization in a ballistic contact could
result in an additional band crossing the Fermi energy, leading to an
abrupt change in the conductance on the order of $e^2/h$.  This has
been theoretically shown to occur in ideal infinite chains of Ni and
Fe~\cite{Velev2005}. However, realistic ferromagnetic contacts made of
transition metals are not expected to be ballistic \cite{Untiedt2004}
and so the interpretation of the conductance jumps in terms of BAMR is
at least questionable.  Indeed, Shi and Ralph~\cite{Shi2007a} have
suggested that these jumps might originate from two-level fluctuations
due to changes in atomic configurations~\cite{Shi2007b}. Thus, there
remain important open questions about AMR in atomic contacts,
concerning the origin of the enhanced amplitude, the anomalous angular
dependence, the occurrence of conductance jumps, and the voltage
dependence.
\begin{figure}[b]
\begin{center}
\includegraphics*[width=\columnwidth,clip]{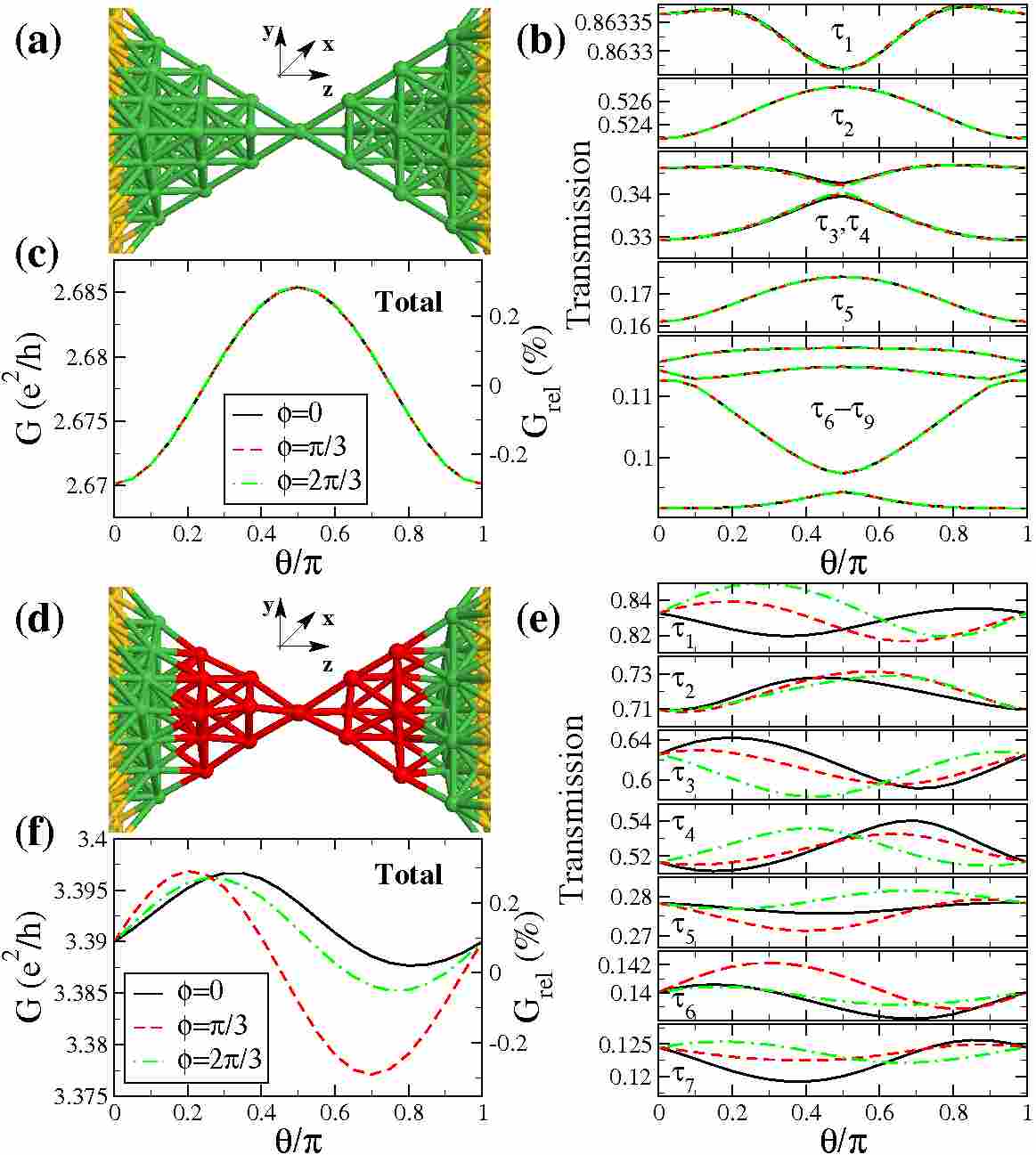}
\caption{\label{AMR_ideal} (Color online) (a) Ideal Ni one-atom
  contact in fcc $[111]$ direction with atoms on lattice positions.
  Green atoms are those in the atomic constriction, yellow ones are
  part of the surfaces used to model the leads. (b,c) Channel
  decomposition and the total linear conductance as a function of
  $\theta$ for different angles $\phi$. The relative conductance is
  defined as $G_{rel}=G(\theta,\phi)/\langle
  G(\theta,\phi)\rangle_{\theta}-1$.  (c) Channel decomposition of
  (b). (d)-(f) Same as (a)-(c), but with the contact distorted by
  randomly shifting the red atoms by up to $5\%$ of the
  nearest-neighbor distance.}
\end{center}
\end{figure}

In this Letter we address these questions using a combination of a
tight-binding (TB) model, molecular dynamics (MD) simulations, and a
simple toy model. Our calculations suggest that the enhancement of the
AMR amplitude and the deviations from the $\cos^2\theta$ law in atomic
contacts stem from spin-orbit coupling (SOC) together with a reduced
symmetry of the contact geometry. We do not find signs of BAMR in
realistic contact geometries. Finally, we find that the conductance of
pure atomic contacts has no voltage dependence on the the scale of
millivolts, but the addition of impurities can lead to a significant
voltage dependence of the AMR signal.

Our description of the AMR in ferromagnetic atomic contacts of the
3$d$ metals (Ni, Co, and Fe) is based on the following non-orthogonal
TB Hamiltonian that includes the SOC:
\begin{equation}
\hat{H} =
\sum_{ij\alpha \beta \sigma \sigma^{\prime}}
\left( h^{(0)}_{i\alpha,j\beta;\sigma}\delta_{\sigma\sigma^\prime}+
h^{(SO)}_{i;\alpha\sigma,\beta\sigma^\prime}\delta_{ij}\right)
\hat c^{\dagger}_{i \alpha\sigma} \hat c_{j \beta\sigma^\prime}.
\end{equation}
Here $i,j$ run over the atomic sites, $\alpha,\beta$ denote the
different atomic orbitals, and $\sigma=\;\uparrow,\downarrow$
the spin. The spin-polarized parameters $h^{(0)}_{i\alpha,j\beta;\sigma}$
and the spin-independent overlaps $s_{i\alpha,j\beta}$ are taken from
the parametrization of Refs.\ \cite{Mehl1996,Bacalis2001},
which is known to accurately reproduce the band structure and total
energy of bulk ferromagnetic materials and has been successfully
applied to the description of ferromagnetic atomic
contacts~\cite{Haefner2008,Pauly2006}. The atomic basis is formed by 9
orbitals ($3d,4s,4p$), which give rise to the main bands around
the Fermi energy in Fe, Co, and Ni. The matrix elements
of the intra-atomic SOC are given by
$h^{(SO)}_{i;\alpha\sigma,\beta\sigma^\prime} = \xi_i \langle
i\alpha\sigma \vert \vec{\bf l} \cdot \vec{\bf s} \vert
i\beta\sigma^\prime\rangle$, where $\vec{\bf l}$ and
$\vec{\bf s}$
are the orbital and spin angular momentum operators. These matrix
elements depend on the spin quantization axis (with polar angle
$\theta$ and azimuthal angle $\phi$), which we rotate
uniformly to simulate the AMR experiments with saturated magnetization.
We use the SOC constant $\xi_i=70$ meV, which is the experimental
value for Ni.  We stress that, while this type of model is not
expected to be as accurate as \emph{ab initio} counterparts, it
captures the essential physics. It also has the advantage that it
allows us to analyze phenomena like BAMR in realistic contact
geometries, which is currently outside the scope of \emph{ab initio}
methods.

\begin{figure*}[t]
\begin{center}
\includegraphics*[width=0.9\textwidth,clip]{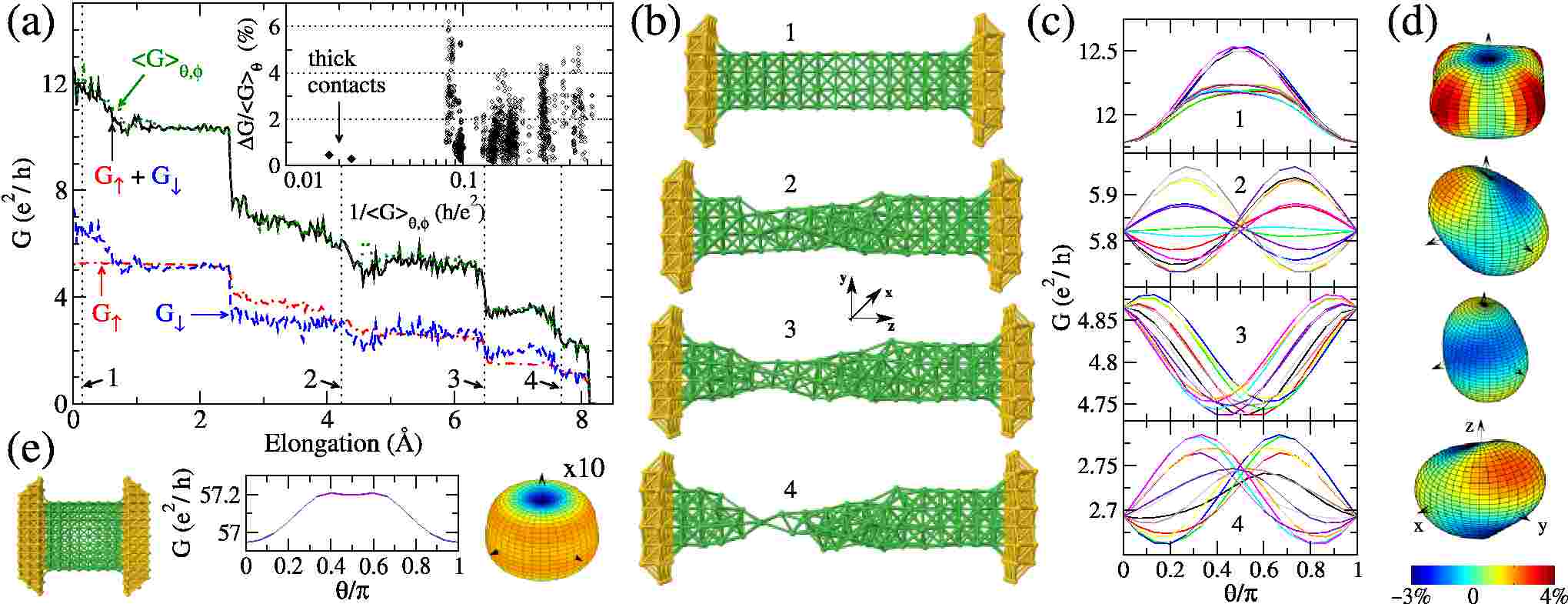}
\caption{\label{md-contacts} (Color online) Contact evolution of a Ni
  junction grown in the fcc $[001]$ direction as obtained from the
  simulations of Ref.\ \cite{Pauly2006}. (a) Spin-projected,
  $G_{\uparrow,\downarrow}$, and total conductance in the absence of
  SOC and total conductance averaged over $\theta,\phi$ in the
  presence of SOC.  Vertical lines correspond to the contact
  geometries in (b).  Inset: relative AMR amplitude
  $\Delta G/\langle G\rangle_\theta
  =(G_{\text{max},\theta}(\phi)-G_{\text{min},\theta}(\phi))/ \langle
  G(\theta,\phi)\rangle_\theta$ vs.\ inverse averaged conductance.
  (c) Conductance vs.\ $\theta$ for the geometries in (b) and with
  $\phi$ in steps of $\pi/6$. (d) Relative AMR
  [$G(\theta,\phi)/\langle G(\theta,\phi)\rangle_{\theta,\phi}-1$] in
  $\%$ on a ``Bloch sphere''. (e) Same as (b)-(d) for a thick regular
  geometry with $324$ atoms with the relative AMR values multiplied by
  $10$ for visibility.}
\end{center}
\end{figure*}

In order to describe the electronic transport we use the
nonequilibrium Green's function formalism
\cite{Additional_material}. Briefly, the atomic contacts are divided
into three parts, a central region $C$ containing the constriction and
the left/right ($L$/$R$) leads, which we model as infinite
surfaces. The retarded Green's functions of the central part read
\begin{equation}
{\bf G}_{CC} = \left[ \varepsilon {\bf S}_{CC}
- {\bf H}^{(0)}_{CC} - {\bf H}^{(SO)}_{CC}
- {\bf \Sigma}_{L} - {\bf \Sigma}_{R} \right]^{-1} ,
\end{equation}
where ${\bf \Sigma }_{X} = {\bf t}_{CX}{\bf g}_{XX}{\bf
t}_{CX}^\dagger$ are the lead self-energies ($X=L,R$).
Here, ${\bf t}_{CX}=\varepsilon {\bf S}_{CX}- {\bf H}^{(0)}_{CX}$,
with ${\bf H}^{(0)}_{CX}$, ${\bf S}_{CX}$ the hopping elements and
overlaps between the $C$ region and the lead $X$, and ${\bf g}_{XX}$
is a surface Green's function. In general ${\bf H}^{(0)}$ and ${\bf
\Sigma}_{L,R}$ depend on the bias voltage $V$. The $V$-dependent
transmission matrix is ${\bf t}(\varepsilon,V) = {\bf
\Gamma}^{1/2}_L {\bf G}_{CC} {\bf \Gamma}^{1/2}_R$, where ${\bf
\Gamma}_X = i({\bf
\Sigma}_{X} - {\bf \Sigma}_{X}^\dagger)$. The current
then adopts the standard Landauer-B\"uttiker-like form
\begin{equation}
I(V) = \frac{e}{h} \int^{\infty}_{-\infty} d\varepsilon \;
\tau(\varepsilon,V)
\left[ f_L(\varepsilon,V) - f_R(\varepsilon,V) \right] ,
\label{current_expression}
\end{equation}
where $f_{L,R}$ are the Fermi functions and
$\tau(\varepsilon,V)=\mbox{Tr} \left[ {\bf t}^{\dagger} {\bf t} \right]$
is the transmission function. The low-temperature linear conductance
can be written as $G= (e^2/h) \sum_n \tau_n$, where $\tau_n$ are the
transmission coefficients, i.e.\ the eigenvalues of ${\bf
t}^\dagger{\bf t}$ at Fermi energy $\varepsilon_F$.

In the calculations presented here we have neglected the SOC in the
leads. We have checked that this only introduces a small change in the
contact resistance that does not alter the conclusions but reduces the
computation time enormously. On the other hand, we determine
self-consistently the on-site energies of the atoms in the
constriction by imposing the local charge neutrality that metallic
elements should exhibit.

We now apply this method to calculate the conductance of Ni atomic
contacts. First we consider an ideal geometry with the atoms kept
fixed on fcc lattice positions and forming pyramid-like tips in the
$[111]$ direction that end in a common central atom
[Fig.\ \ref{AMR_ideal}(a)]. In Figs.\ \ref{AMR_ideal}(b,c) the
conductance and its channel decomposition as function of $\theta$ for
several values of $\phi$ are shown.  Surprisingly, the conductance of
this one-atom contact exhibits the bulk-like AMR with a $\cos^2\theta$
dependence (minimum at $\theta=0$), an amplitude of $0.5\%$ and
practically no dependence on $\phi$. In fact, the individual channels
show a more complicated dependence on $\theta$, and the amplitude of
variation for one channel can be bigger than that of the total
conductance, but in the latter these features cancel and the
$\cos^2\theta$ dependence is recovered.

It seems obvious that the cancellation is related to the high symmetry
of the ideal geometry. To test this idea we have distorted the
contact by shifting randomly the atomic positions by up to $5\%$ of the
nearest-neighbor distance [Fig.\ \ref{AMR_ideal}(d)]. As seen in
Fig.\ \ref{AMR_ideal}(e), the individual channels now show roughly the
same amplitude of variation with $\theta$ as in the ideal contact, but
due to the disorder they exhibit a more complex $\theta$ dependence,
and a strong dependence on $\phi$. As a consequence, the contributions
of the channels no longer cancel out and the AMR can have a different
amplitude, with the conductance extrema shifted in $\theta$ and with a
strong dependence on $\phi$ [Fig.\ \ref{AMR_ideal}(f)].  This example
illustrates that the origin of the anomalous angular dependence and
amplitude of the AMR in atomic contacts can be simply the reduced
symmetry of these junctions together with the fact that the
conductance is mainly determined by a few atoms in the narrowest part
of the constrictions.  We want to point out that we have found similar
results for Co and Fe atomic contacts, which confirms this conclusion
\cite{Additional_material}.

Since the geometry plays such a prominent role in the AMR, it is
important to determine the geometries that can be realized in an
actual experiment. For this purpose, we have carried out classical MD
simulations of the formation of Ni atomic contacts, following
Ref.\ \cite{Pauly2006}.  An example of the contact evolution is
shown in Fig.\ \ref{md-contacts}. Here, we start with an ideal Ni bar
containing 112 atoms on lattice sites in fcc $[001]$ direction. The
bar is attached to rigid surfaces that are separated in a stepwise
manner, simulating the elongation process of a break junction.  In
Fig.\ \ref{md-contacts}(a) we show the evolution of the spin-projected
and total conductance during elongation in the absence of SOC. Adding
SOC introduces only a small change in the averaged total conductance.
As usual, the sudden atomic rearrangements are reflected as steps in
the conductance~\cite{Agrait2003}. The vertical lines refer to the
geometries of Fig.\ \ref{md-contacts}(b) obtained during the
elongation.  For them we have computed the dependence of the
conductance on $\theta$ and $\phi$, and the results are shown in
Fig.\ \ref{md-contacts}(c).  For contact 1, which is just an elastic
deformation of the ideal contact, the conductance has two types of
behavior depending on $\phi$: one is $\cos^2\theta$-like, while the
other is clearly more complex.  In order to visualize the overall
angular dependence, we show the relative AMR on a ``Bloch sphere''
[Fig.\ \ref{md-contacts}(d)]. The contact 1 and hence its AMR have an
approximate four-fold symmetry. When deformations emerge in the
contact, the angular dependence becomes irregular and strongly
dependent on $\phi$.  For example, for contact 2 there is a strong
variation of AMR with $\phi$, and depending on its value, the AMR
amplitude can be almost one order of magnitude larger than in the
bulk limit or cancel almost entirely. As the contact evolution
proceeds, the AMR has an amplitude of around $2\%$. In almost all
cases, the conductance is not only shifted in $\theta$, but it also
has a more complicated behavior than just $\cos^2\theta$. Approaching
the tunnel regime ($G<0.1e^2/h$), we do not observe a further
increase of the AMR amplitude, contrary to
experiments~\cite{Bolotin2006}.  One reason may be that the isolated
tip atoms in tunneling regime exhibit a finite orbital
moment~\cite{Desjonqueres2007} (not considered here), which may lead
to a local deviation of the spin-quantization axis from the field
direction and an additional increase in resistance. Finally, in the
limit of thick contacts, we recover bulk behavior with an amplitude of
$0.45\%$, as shown in Fig.\ \ref{md-contacts}(e). The statistical
analysis of all contact geometries shows an increase of AMR to $2\%$
on average in the last steps before breaking, see inset of
Fig~\ref{md-contacts}(a). This confirms that the lack of
symmetry in atomic contacts gives rise to the enhancement of
the AMR signal.

\begin{figure}[t]
\begin{center}
   \includegraphics*[width=\columnwidth,clip]{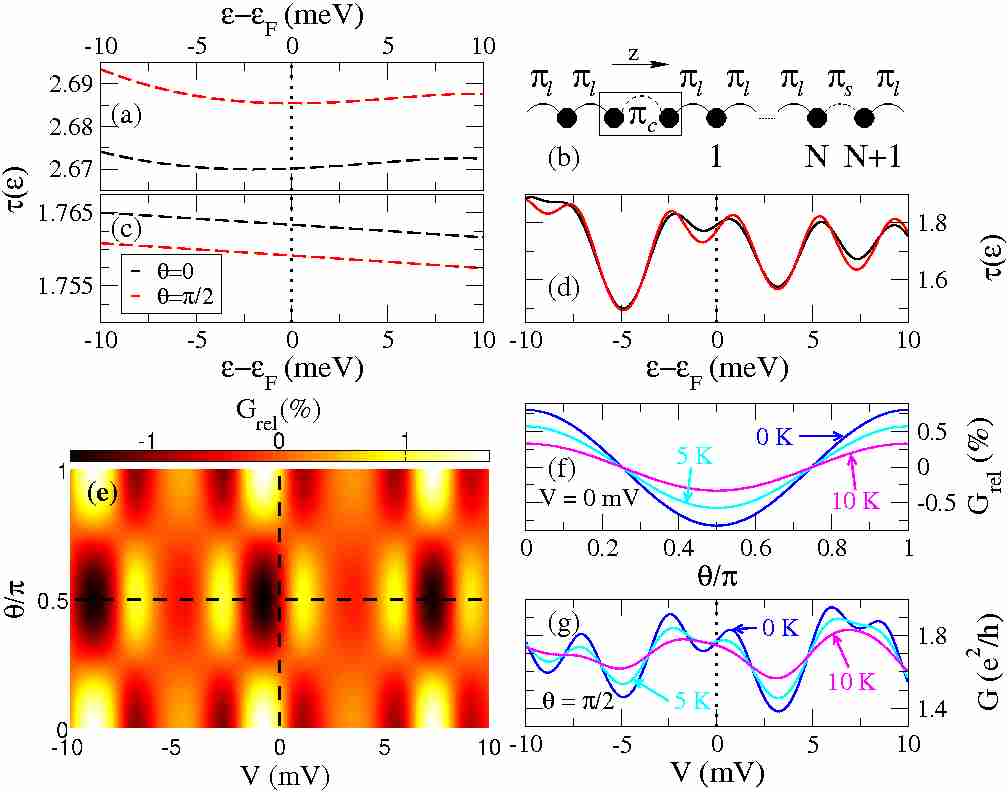}
   \caption{\label{Impurities} (Color online) (a) $\tau(\varepsilon,V=0)$
   for the contact of Fig.\ \ref{AMR_ideal}(a) with $\theta=0$,$\pi/2$.
   (b) Sketch of the chain model, (c) corresponding transmission for
   $\theta=0$,$\pi/2$ in the absence of impurities, (d)
   $\tau(\varepsilon,V=0)$ with an impurity $N=751$ sites from the
   scattering region, and (e) relative nonlinear conductance
   $G_{rel}(\theta,V)=G(\theta,V)/\langle G(\theta,V)\rangle_\theta-1$.
   (f) Relative linear conductance vs.\ $\theta$ and (g) voltage
   dependence of nonlinear conductance at $\theta=\pi/2$ for indicated
   temperatures. Results at $0$ K in (f,g) correspond to the dashed
   lines in (e).}
\end{center}
\end{figure}

We have not found indications of tip resonances, which are present
in ideal one-dimensional geometries~\cite{Burton2007,Jacob2008} and
which were suggested as the origin of the experimental
findings~\cite{Burton2007}. For example, the transmission for the
contact of Fig.\ \ref{AMR_ideal}(a) has almost no structure around
$\varepsilon_F$ on the scale of millielectronvolts, as shown in
Fig.\ \ref{Impurities}(a). We thus believe that the voltage and
temperature dependences reported in
Refs.\ \cite{Bolotin2006,Shi2007b} are indeed associated with
impurities close to the constriction, as reported earlier for
non-magnetic junctions~\cite{Ludoph1999,Untiedt2000}.
Following Ref.\ \cite{Untiedt2000}, one may estimate that the
reported \cite{Bolotin2006} voltage period of a few millivolts can
stem from impurities located hundreds of nanometers away from the
contact. Such length scales cannot be modeled realistically, but we
have developed a toy model to support the idea. As represented
schematically in Fig.\ \ref{Impurities}(b), we model the contact with
two semi-infinite linear chains connected at the tips, and describe
the system with a nearest-neighbor TB Hamiltonian.  For simplicity we
only include the $d_{xy}$ and $d_{x^2-y^2}$ orbitals (the $E_{2}$
doublet), with a spin-splitting of $\Delta=0.5$ eV.  In the absence of
SOC there are four independent equivalent chains with onsite energies
$\pm \Delta/2$, hopping integrals $\pi_l=1.0$ eV in the leads, and
$\pi_c=0.4$ eV between the tip atoms.  The SOC that couples the chains
is restricted to the two tip atoms, with a coupling constant $\xi=0.2$
eV.  For further details, see Ref.\ \cite{Additional_material}.
As we show in Fig.\ \ref{Impurities}(c), there is no structure in
$\tau(\varepsilon,V=0)$ and the AMR is small due the high symmetry of
the geometry. In this example $\varepsilon_F=0.42$ eV, which makes the
sign of the AMR opposite to that in Fig.\ \ref{Impurities}(a).  We
note that the results of Fig.\ \ref{AMR_ideal}(b) are
qualitatively reproduced \cite{Additional_material}.

Things change drastically when an impurity is introduced in one of the
chains. We model it by using slightly reduced hoppings $\pi_s=0.8$ eV
between two atoms $N=751$ sites away from the constriction in the
right lead.  Figure \ref{Impurities}(d) shows the resulting
transmission functions for $\theta=0,\pi/2$. In addition to the
expected oscillations in energy, there appears a modulation depending
on $\theta$ that stems from the interference with the impurity.  When
a finite voltage $V$ is applied over the tip atoms, the nonlinear
conductance $G(V)=dI/dV$ exhibits oscillations as a function of both
voltage and the angle $\theta$. The relative variations shown in
Fig.\ \ref{Impurities}(e) bear a striking resemblance to the
experimental results of Ref.\ \cite{Bolotin2006}.  Finally, we
show in Figs.\ \ref{Impurities}(f,g) the temperature dependence of both
the linear and non-linear conductance. The effect of temperature is
to smooth the $0$ K characteristics, again compatible
with the experimental observations~\cite{Shi2007b}.

In summary we have shown that the anomalous magnitude and angular
dependence of the AMR in ferromagnetic atomic-sized contacts can be
explained naturally in terms of the reduced symmetry of the atomic
junction geometries. We predict a strong anisotropy of the conduction
channels, but have not found any signature of BAMR. We have presented
a simple model which illustrates that the pronounced voltage and
temperature dependence found in some experiments may originate from
the presence of impurities close to the constriction.

We thank M.\ Dreher, G.\ Sch\"on, A.\ Levy Yeyati, N.\ Agra\"{\i}t, M.\
Viret, and E.\ Scheer for helpful discussions. We acknowledge support
from the Helmholtz Gemeinschaft (Contract No.\ VH-NG-029), and the DFG
within the CFN.  MH also acknowledges support by the KHYS.


\end{document}